\documentstyle[12pt]{article}

\renewcommand{\thefootnote}{\fnsymbol{footnote}}
\newcommand{\EQ}{\begin{equation}}
\newcommand{\EN}{\end{equation}}
\newcommand{\bea}{\begin{eqnarray}}
\newcommand{\ena}{\end{eqnarray}}
\newcommand{\vs}[1]{\vspace{#1 mm}}

\newcommand{\pa}{\partial}

\newcommand{\uda}{\nearrow \kern-1em \searrow}

\begin{document}

\topmargin 0pt
\oddsidemargin 5mm

\begin{titlepage}
\setcounter{page}{0}
\begin{flushright}
April, 1998\\
OU-HET  291\\
hep-ph/9804245 \\
\end{flushright}

\vs{12}
\begin{center}
{\Large  $e^- e^-$ Collisions Mediated by Composite Neutrinos}
\vs{15}

{\large Robert Kinyua\footnote{e-mail address:kubai@phys.wani.osaka-u.ac.jp} 
and 
Eiichi Takasugi\footnote{e-mail address: takasugi@phys.wani.osaka-u.ac.jp}}\\
\vs{8}
{\em Department of Physics,\\
Osaka University Toyonaka 560} \\
\end{center}
\vs{6}

\centerline{{\bf{Abstract}}}
The lepton number violating process, $e^- e^- \to W^- W^-$ has been 
widely discussed in the Majorana neutrino exchange mechanism.  
Here, we discuss this process in a composite neutrino model where  
excited Majorana neutrinos are exchanged. 
We found several qualitatively different features from the neutrino 
exchange case: (1) The longitudinally polarized 
$W$'s are not produced, (2) the neutrinoless double beta decay does not 
constrain much and a much larger cross section is expected, and (3) 
CP violating phases may be explored because all excited neutrinos are 
heavy so that large mixings among them are expected. 
\end{titlepage}
\newpage
\renewcommand{\thefootnote}{\arabic{footnote}}
\setcounter{footnote}{0}
\section{Introduction}

Rizzo[1] was the first to analyze the $e^- e^- \to W^- W^-$ scattering  
mediated by Majorana neutrinos and a triplet Higgs 
boson in both an extended standard model and the left-right 
symmetric model. The interest lies in the fact that 
if this process is observed, 
one can say that at least the electron neutrino is a massive Majorana 
neutrino by using a similar argument to the one made for 
the neutrinoless double beta $(\beta\beta)_{0\nu}$ 
decay[2]. Later, many authors[3],[4] have paid much attention to this process 
mediated by  heavy neutrinos, and the 
feasibility of observing it by the future TeV linear colliders was discussed. 
However, Belanger et al.[4] have argued that it may not be possible 
to observe this scattering with the next linear collider (NLC) 
of $\sqrt s = 1$TeV and 
an energy of at least 4TeV is needed to observe this. This unpleasant 
result is due to the constraint from neutrinoless double beta decay. 

In this paper, we analyze this process in a composite model. In this 
scenario, the scattering occurs by the exchange of excited neutrinos. 
We found qualitatively different features from  
the Majorana neutrino exchange case: (1) 
The longitudinally polarized $W$'s are not produced in this mechanism, 
while the production of longitudinally polarized $W$'s is the dominant mode 
in the neutrino exchange case. (2) A relatively large cross section 
is expected in comparison with the tiny one in the neutrino 
exchange case where the 
the neutrinoless double beta $(\beta\beta)_{0\nu}$ decay gives 
a severe constraint[4]. The $(\beta\beta)_{0\nu}$ 
decay is considered in a composite neutrino model by Takasugi[5], and 
Panella et al.[6], where the $(\beta\beta)_{0\nu}$ decay takes place by the 
exchange of excited neutrinos.  In this paper, we show that 
 the constraint form the $(\beta\beta)_{0\nu}$ decay does give only  a 
mild constraint on the cross section.   
(3) Large mixings of electron to heavy excited neutrinos are expected 
because all excited neutrinos are heavy, whereas for the heavy neutrino 
mediated case, mixings of electron to heavy neutrinos are tiny.  
Thus, there  could be  
a good chance to explore the CP violation phases in Majorana 
neutrino system. 
 
In Sec. 2, we explain the $(\beta\beta)_{0\nu}$ decay constraint and 
give the cross section formula. 
 In Sec. 3, the numerical analysis is given. The CP violation in 
 heavy neutrino system is discussed in Sec.4. 
Summary is presented in Sec.5.

\section{The cross section}

\noindent
A. Interaction
 
Excited neutrinos  couple to the ground state leptons by the dimension five 
magnetic coupling[7].  This interaction is expressed as[8]
\bea
L_{int}=g\frac{\lambda_W }{ m_{\nu^*}}\bar e 
\sigma^{\mu\nu}(\eta_L^* R+\eta_R^* L)\nu^*_e \pa_\mu W_\nu^-  + h.c. ,
\ena
where $\nu^*$ is a heavy excited electron neutrino, $L=(1-\gamma_5)/2$, 
$R=(1+\gamma_5)/2$, and $m_{\nu*}$ is the mass dimension which is of 
order of the mass of $\nu^*$, i.e., $m_*$. This interaction may  arise from 
an $SU(2)\times U(1)$-invariant higher-dimensional interaction[7],[8]. 
Normalization parameters $\eta_L$ and  $\eta_R$ are given by  $(\eta_L, \eta_R)
  =(1,0)$  or (0,1) to respect the chirality conservation. 
We consider the mixing among excited neutrino. The excited electron neutrino 
$\nu_{e}^*$ may be expressed by a superposition of mass eigenstate 
Majorana neutrinos $N_j^*$ with the mass $m_j$ as 
\bea
\nu^*_e = \sum_{j} U_{ej} N_j^*\;.
\ena 
Extensive search for $\nu^*$ have been made by accelerator experiments[9] 
and it has been found that $m_{\nu*} > 91$ GeV by assuming 
that  $\lambda_Z > 1$, which is the coupling for 
$\nu^* \to \nu Z$ decay similarly defined  to  $\lambda_W $. 
Hereafter, we assume that the excited neutrino mass is much larger than 
the W boson mass, i.e., $m_j>>m_W$.   

Neutrinoless double beta decay mediated by excited neutrinos 
in composite models has been examined  by 
Takasugi[5] and Panella et al.[6]. By comparing the theoretical calculation 
with the Heidelberg-Moscow data[10] 
for ${}^{76}$Ge decay, $T_{1/2}^{0\nu}>1.2\cdot 10^{25}$yr(90$\%$C.L.), the 
following constraint on the coupling parameter is given  
\bea
\left( \frac{\lambda_W }{\hat m_{\nu*}}
\right)^2
\left |\sum_j U_{ej}^2 \left (\frac{\hat m_j+2}{(\hat m_j +1)^2}
 -\frac{0.129}{\hat m_j}\right) \right | <1.4\times 10^{-2}\;,
\ena
where the hatted quantities are dimensionless ones scaled by $m_W$;  
$\hat m_{\nu*} \equiv m_{\nu*}/m_W$ and  $\hat m_j \equiv m_j/m_W$.  
Fig.1 shows the upperlimit of $\lambda_W /\hat m_{\nu*}$ as a function 
of $m_1$ which is denoted by $m_*$, by assuming $U_{e1}=\delta_{j1}$. 
This situation is reasonable because  all excited 
neutrinos are heavy so that some mixings $U_{ej}$ must be large. From this 
figure, we see that $\lambda_W /\hat m_{\nu*}<0.33$ for $m_*>0.5$TeV 
and the bound becomes less severe as the excited neutrino mass increases. 
 
\noindent
B. Cross section

From the interaction in Eq.(1), the invariant amplitude for the 
$e^- e^- \to W^- W^-$ occuring via the t- and u-channel exchange of 
excited neutrinos is given by 
\bea
m_{fi}&=&\left ( g\frac{\lambda_W }{ m_{\nu^*}}\right)^2 \sum_j m_j U_{ej}^2 
\left\{ \frac{k_{1\mu}k_{2\rho}\epsilon_{\nu}(k_1) 
  \epsilon_{\sigma}(k_2)}{(k_1-p_1)^2-m_j^*}+ 
   \frac{k_{1\mu}k_{2\rho}\epsilon_{\nu}(k_1) 
   \epsilon_{\sigma}(k_2)}{(k_1-p_1)^2-m_j^*}
\right\}\cr
&&\times {\bar {u^c}}(p_1)\sigma^{\mu\nu}\sigma^{\rho\sigma}
 (\eta_L^2L+\eta_R^2R)u(p_2)
\ena
where $u^C$ is the charge conjugation of $u$. From the structure of 
the above amplitude, one finds that 
the longitudinally polarized $W$ ($\epsilon_{\mu}(k) \simeq k_{\mu}$) 
can not be produced in the excited 
neutrino exchange mechanism.  
 In other words, the $\alpha$ part of 
\bea
\sum_{\lambda} \epsilon_{\mu}^{(\lambda)}(k_1)\epsilon_{\mu'}^{(\lambda)*}(k_1)
=-g_{\mu\mu'}+\alpha k_{1\mu}k_{1\mu'}
\ena
does not contribute to the spin sum of $|m_{fi}|^2$. 
In contrast, the longitudinally polarized $W$ production is the dominant 
contribution in the heavy neutrino exchange mechanism for $\sqrt s >> m_W$.

The cross section is  calculated from Eq.(4) by neglecting the electron mass 
and the differential cross section  for the unpolarized electron 
beam is given by using the 
invariant variables, $s$, $t$, and $u$ as 
\bea
\frac{d\sigma}{d\cos \theta}&=&
 \frac{g^4}{64\pi m_W^2 }\left ( \frac{\lambda_W }
 {\hat m_{\nu^*}}\right)^4 \frac 1{\hat s}
 \sqrt{1- \frac 4{\hat s}}\sum_{i,j} \hat m_i \hat m_j U_{ei}^2 U_{ej}^{2*}
 f_{ij}\;,
\ena 
where 
\bea
f_{ij}&=& \frac1{(\hat t -\hat m_i^2)(\hat t -\hat m_j^2)}
\left [(\hat s-2)(\hat t-1)^2-(\hat t-1)(\hat u-1)+\frac{\hat s}{4}\right ]\cr
&&\; + \frac 1{(\hat u -\hat m_i^2)(\hat u -\hat m_j^2)}
\left [(\hat s-2)(\hat u-1)^2-(\hat t-1)(\hat u-1)+\frac{\hat s}{4}\right]\cr
&&\;+ \frac12 \left( \frac 1{(\hat t-\hat m_i^2)(\hat u -\hat m_j^2)} 
+\frac 1{(\hat u -\hat m_i^2)(\hat t -\hat m_j^2)}\right)\cr
&&\times \left (\hat s (\hat s-2)^2-(\hat s-2)[(\hat t-1)^2+(\hat u-1)^2]
 +2(\hat t-1)(\hat u-1)-\frac 32 \hat s\right )\;,
\ena
and all hatted-variables are dimensionless ones scaled by $W$ mass as 
$\hat m_{\nu*}=m_{\nu*}/m_W$, $\hat s=s/m_W^2$, $\hat m_j=m_j/m_W$ and so on. 
The invariant variables satisfy the relation $\hat s+\hat t+\hat u = 2$. 

For $\hat s >>1$ ($s>>m_W^2$), the cross section takes the simpler forms   as 
\bea
\frac{d\sigma}{d\cos \theta}&=&
 \frac{g^4}{64\pi m_W^2}\left 
 ( \frac{\lambda_W }{\hat m_{\nu^*}}\right)^4
 \left | \sum_{i} \hat m_j  U_{ej}^2 
  \left(\frac{\hat t}{(\hat t -\hat m_j^2)}+\frac{\hat u}{(\hat u -\hat m_j^2)}
   \right)\right|^2\cr
   &&\to \frac{g^4}{16\pi m_W^2}\left 
 ( \frac{\lambda_W }{\hat m_{\nu^*}}\right)^4
 \left | \sum_{j} \hat m_j  U_{ej}^2 \right|^2\quad {\rm for}\quad  s>>m_j^2\;,
 \cr
 &&\to \frac{g^4}{64\pi m_W^2}\left 
 ( \frac{\lambda_W }{\hat m_{\nu^*}}\right)^4 
 \left | \sum_{j} \frac {U_{ej}^2}{\hat m_j} \right|^2 \hat s^2 \quad 
 {\rm for}\quad  s<<m_j^2\;. 
\ena 
The above cross section formula is similar to the one for the neutrino 
mediated case and  violates the unitarity 
bound for $s \to \infty$. This difficulty may be avoided by assuming 
that masses and mixing angles satisfy $\sum m_j U_{ej}^2 =0$, similar 
to the heavy neutrino exchange case[3],[4].  
If  polarized beams are used such as in $e_{L} e_{L}$ scattering which occurs 
for $(\eta_L,\eta_R)=(1,0)$ or $e_{R} e_R$ for $(\eta_L,\eta_R)=(0,1)$, 
the cross section should be multiplied by 4.
 
\section{The expected cross section}

Firstly, we examine the $(\beta\beta)_{0\nu}$ decay constraint on the 
cross section for $m_W<<\sqrt s<<m_j$. In this region of $\sqrt s$, 
the $(\beta\beta)_{0\nu}$ constraint in Eq.(3) takes a simpler form as  
$({\lambda_W }/{\hat m_{\nu*}})^2|\sum_j {U_{ej}^2}/{  m_j} | 
<1.4\times 10^{-2}$. Then, by combining this constraint and the 
formula in Eq.(8) for $s<<m_j^2$, we 
 find that the upper bound of the cross section 
for the unpolarized beam 
\bea
\frac{d\sigma}{d\cos \theta}<11 
\left(\frac{\sqrt{ s}}{m_W}\right)^4 \quad {\rm fb}\;. 
\ena 

Suppose  that the NLC integrated luminosity is  about 80fb${}^{-1}$ at 
$\sqrt s =$1TeV[11]. In contrast, the upper limit  of the differential 
cross section 
given in Eq.(9) is about $3\times 10^5$fb so that we can say that 
the  cross section is essentially not constrained by the neutrinoless 
double beta decay in comparison with the expected luminosity. 
Therefore, we have to set a more realistic condition. A reasonable 
guess is that the relative coupling strength is of order one, 
i.e., $\lambda_W=1$ and examine the dependence of the cross 
section on masses of excited neutrinos and the composite scale. 
In the reminder of this section, we assume for simplicity that 
one of the mixing is dominant, i.e.,  $U_{ej}=\delta_{j1}$. We denote 
the mass $m_1$ as $m_*$ and express the composite scale as $\Lambda_C 
\equiv m_{\nu*}/\sqrt 2$. 

In Fig.2, the center of mass energy dependence of the cross section 
for various excited neutrino mass $m_*$ with $\Lambda_C$=1TeV and 
$\lambda_W=1$. For energies less than $m_*$, the cross section 
is a decreasing function of $m_*$. If the energy exceeds $m_*$, 
the cross section becomes an increasing function of $m_*$. 
Therefore, the cross over between two curves for 
$m_*=1$ and 2TeV  around $\sqrt s=1.9$TeV appears. These 
behaviors can be seen analytically from the cross section formula 
in Eq.(8). It should be noted however that these behavioures 
are valid for energies which do not exceed the excited 
neutrino mass much.

Next we consider how far we can explore  $\Lambda_C$ and 
$m_*$ by using NLC with the luminosity about 80fb${}^{-1}$. 
We assume that the minimum cross section needed to be detected 
is $\sigma$=0.1fb, i.e., 
eight events  in a year. Then,  we estimated the size  
of  $\Lambda_C/\lambda_W$ and $m_*$ which can be explored.  The result is 
shown in Fig.3 where the lower region from the curve is the region which 
corresponds to $\sigma>$0.1fb.  Roughly speaking, we can explore about 
20TeV scale of $m_*$  for $\Lambda_C$ of about a few TeV by NLC.
 
\section{CP violation}

Once the $e^-+e^- \to W^-+W^-$ scattering is observed and the cross 
section is much greater than  $5\times 10^{-3}$fb  at $\sqrt s=$1TeV 
which is the upper bound for the heavy neutrino exchange case[4], 
the production mechanism would be the excited neutrino exchange in a 
composite neutrino scenario.  More decisively, 
if the produced $W$'s turn out to have only the transverse polarization, 
the composite neutrino scenario is the only candidate at present. 
If the scattering is found,  CP violation will become one of the urgent 
subjects to examine. In below, we shall discuss how  CP violation phases 
get in the cross section formula. The following discussion is valid 
both for heavy neutrino scenario and also for composite neutrino scenario. 

In general, as discussed by Bilenky, Hosek and Petcov[12], and by Doi, 
Kotani, Nishiura, Okuda and Takasugi[13], 
there are extra  CP violating phases in Majorana neutrino system and 
they  contribute to the lepton number violation process, in addition to 
the KM like CP violation phase which is intrinsic to Dirac system. 
In particular, we can parameterize  mixing matrix elements 
for three generation as[13]
\bea
U_{e1}=c_1 e^{i\alpha}\;,\quad U_{e2}=-s_1c_3 e^{i(\alpha+\beta)}\;,
\quad U_{e3}=-s_1s_3 e^{i(\alpha+\gamma)}\;, 
\ena
where $s_i=\sin \theta_i$ and $c_i=\cos \theta_i$, $\theta_i$ 
is the rotational angle around the $i$th axis of flavor space, and 
$\beta$ and $\gamma$ are CP violation phases. 
Then, the cross section behaves as  
\bea
\frac{d\sigma}{d\cos \theta}&\propto &
\hat m_1^2c_1^4 f_{11}+\hat m_2^2s_1^4c_3^4 f_{22}+\hat m_3^4s_1^4s_3^4 f_{33}
+2\hat m_1^2\hat m_2^2 s_1^2c_1^2c_3^2\cos 2\beta f_{12}\cr
&&+2\hat m_1^2\hat m_3^2 s_1^2c_1^2s_3^2\cos 2\gamma f_{13}
+2\hat m_2^2\hat m_3^2 s_1^4s_3^2c_3^2\cos 2(\beta-\gamma) f_{23}\;,
\ena
where $f_{ij}$ is defined in Eq.(7). Thus, two CP violating phases 
$\beta$ and $\gamma$ appear. Among two phases, one is 
the phase intrinsic to a Dirac system and the other is the one intrinsic 
to a Majorana system. The above is a general form which corresponds to a 
general neutrino mass matrix.   

Here, we face to a problem of a constant behavior of the cross section 
 as $\sqrt s \to \infty$, similar to neutrino exchange case. 
The constant behavior means the violation of the unitarity. 
There must be some mechanism to restore the unitarity. 
For composite model, it is out of our scope to introduce  
some new interaction to remedy this defect and thus we assume that 
the excited neutrino mass matrix is arranged such that 
\bea
 m_{\nu e\nu e}=(UDU^T)_{\nu e\nu e}=\sum_{j}m_j U_{ej}^2=0\;,
\ena
where $D$ is a diagonal mass matrix. This condition is  
similar to the heavy neutrino scenario.   Because of this constraint, 
only one CP violation phase appears for three generation case and 
no CP violation phase appears for two generation case. 
This may be seen by rewriting the constraint as 
\bea
m_1c_1^2+m_2s_1^2c_3^2e^{2i\beta} +m_3s_1^2s_3^2 e^{2i\gamma}=0\;.
\ena
For three generation case, 
$\beta$ and $\gamma$ are not independent each other. Two generation mixing 
may be derived by setting  $s_3=0$. Then, Eq.(14) forces $\beta=0$ or $\pi$ 
so that there appears  no CP violation. 
This can be seen explicitly by using the mass 
matrix. We consider a symmetric  matrix which respect the condition (13). 
We find that all phases can be absorbed by the phase matrix $P$ as
\bea
\left(\matrix{0&ae^{i\alpha}\cr ae^{i\alpha}&be^{i(\alpha+\beta)}\cr}\right )
=P\left(\matrix{0&a\cr a &b}\right)P\;,
\ena
where  $P=$diag$(e^{i(\alpha-\beta)/2},
e^{i(\alpha+\beta)/2})$. Then, the mixing matrix $U$ can be 
expressed as $U=P^{\dagger}O$, where $O$ is a orthogonal matrix. 
Then, no CP violation phase appears in the $e^-+e^-\to W^-+W^-$ reaction. 

\section{Summary}

We have discussed the  $e^-+e^-\to W^-+W^-$ reaction in the composite neutrino 
scheme. In the composite model, the excited neutrinos couples to the 
ground state electron and $W$. By using the gauge invariant interaction 
of dimension five, we analyzed this process by exchanging 
the excited neutrino. We found various new features which are 
different from the neutrino exchange case as discussed in the introduction. 
Since the $(\beta\beta)_{0\nu}$ decay does not constrain the cross section, 
it is worthwhile and interesting to explore this process in NLC 
with $\sqrt s =1$TeV and the luminosity of 80fb${}^{-1}$. If this 
reaction is observed, it is likely that neutrinos are composite and 
that there exist excited states. The decisive confirmation can be made 
by observing the polarization of $W$. In the composite scenario,  
longitudinally polarized $W$'s are not produced.

\newpage

\noindent
Figure Captions:

\noindent
Fig.1: The upper bound of the coupling strength of excited neutrinos and the 
ground state electron and $W$ imposed by the neutrinoless double beta decay.   
The area below the curve is the allowed region. 

\vskip 5mm
\noindent
Fig.2: The center of mass energy $\sqrt s$ dependence of the cross section 
for various values of the excited neutrino mass $m_*$ with 
the composite scale $\Lambda_C=$1TeV with $\lambda_W=1$. 

\vskip 5mm  
\noindent
Fig.3: The region of $\Lambda_C/\lambda_W$ and $m_*$ for which  
one can explore, 
where  the cross section is fixed to be 0.1fb. The below the curve 
is the region of parameters where $\sigma >$0.1fb.


\newpage
\begin{thebibliography}{99}
\bibitem{TGR}
T.G. Rizzo,
Phys. Lett. B116 (1982), 23.
\bibitem{UR}
J. Schechter and J. W. F. Valle, Phys. Rev. D25 (1982), 2951.\\
E. Takasugi, Phys. Lett. B149 (1984), 372.
\bibitem{LBN}
D. London, G. B{\'{e}}langer,J.N. Ng,
Phys. Lett. B188 (1987), 155;\\
C.A. Heusch, P. Minkowski,
Nucl. Phys. B416 (1994), 3;\\
see also Proceedings of the electron-electron linear collider workshop, 
Int. Jour. Mod. Phys. A11 (1996),1523.
\bibitem{BBLN}
G. B{\'{e}}langer, F. Boudjema, D. London, H. Nadeau,
Phys. Rev. D53 (1996), 6292.
\bibitem{ET}
E. Takasugi,
Prog. Theor. Phys. 98 (1997), 977.
\bibitem{PCSW}
O. Panella, C. Carimalo, Y.~N. Srivastava and A. Widom,
Phys. Rev. D56 (1997), 5766.
\bibitem{CMS}
N. Cabibbo, L. Maiani and Y. Srivastava,
Phys. Lett.  139B (1984), 459.\\
K. Hagiwara, S. Komamiya and D. Zeppenfeld,
Z. Phys. C29 (1985), 115.
\bibitem{PDG}
Particle Data Group,
Review of Particle Physics, Phys. Rev. D50 (1994), 1173.
\bibitem{MG}
L3 Collaboration,
Phys. Rep. C236 (1993), 1.\\
ALEPH Collaboration,
Phys. Rep. C216 (1992), 253.\\
See references in Particle Data Group.
\bibitem{HM}
H.V. Klapdor-Kleingrothhaus, hep-ex/9802007. 
\bibitem{TWM}
see for example, T.W. Markiewicz, Int. Jour. Mod. Phys. A11 (1996) 1649.
\bibitem{BHP}
S.M. Bilenky, J. Hosek and S.T. Petcov, Phys. Lett. 98B (1980), 495.
\bibitem{DKNOT}
M. Doi, T. Kotani, H. Nishiura, K. Okuda and E. Takasugi, Phys. Lett. 102B 
(1981), 323.
\end{thebibliography}
\end{document}